\documentclass[aps,prl,twocolumn,showpacs]{revtex4}

\usepackage{graphicx}%
\usepackage{dcolumn}
\usepackage{amsmath}

\makeatletter
\def\btt#1{\texttt{\@backslashchar#1}}%
\DeclareRobustCommand\bblash{\btt{\@backslashchar}}%
\makeatother

\begin{document}

\title[Short Title]{Composite quasiparticles and the ``hidden" quantum critical point\\
in the topological transition 
scenario of high-$T_c$ cuprates}
 
 \date{\today}

\begin{abstract}
The quantum interference effects due to the Aharonov-Bohm-type phase factors are studied in 
the layered $t-t'-t_\perp-U-J$ strongly correlated system  relevant for cuprates. 
Casting  Coulomb  interaction 	 in terms
of composite-fermions via the flux attachment facility, we argue that U(1) compact group
instanton  events labelled by a topological winding number
are essential configurations of the phase field dual to the charge.
The impact of these  topological excitations is calculated for the phase diagram
which displays the ``hidden" quantum critical point of a novel type.
\end{abstract}

\author{T. K. Kope\'{c}}
\affiliation{
Institute for Low Temperature and Structure Research,
Polish Academy of Sciences,
POB 1410, 50-950 Wroclaw 2, Poland}
\maketitle

The discovery of high-temperature superconductors (HTSC) and follow--up  studies of
strongly correlated (SC) fermionic systems  reveal that  
an explanation of their   unusual properties appears
unlikely in a way of thinking rooted in an independent electron picture.
 It is widely accepted that the central issue in the high
temperature  superconducting  cuprates 
is physics of the doped Mott insulator \cite{mott}.
There are also  strong indications \cite{qcp}, that much
of their behavior is governed by the proximity to a kind of quantum critical point (QCP).
 However,  in approaches to the  HTSC   one customary concentrates
on the low-energy physics usually discarding the hallmark of SC systems - the high
energy scale given by the  Coulomb interaction $U$ by projecting out  double-occupancy
charge configurations.
A detour from the strict projection program was recently proposed in a form of
the ``gossamer" superconductor \cite{laughlin}
recognizing the role of  correlations among expensive double--occupancy charge
configurations.

Moreover, a  SC electronic
system can have non-trivial
{\it topological} properties which can be described by  gauge fields \cite{frohlich}:
a  phase of the many-body wave function might be arbitrary
but  correlations among the local phases of
its constituents  can bring unusual gauge structures \cite{wilczek}.
Quantum theories with topological properties have raised considerable
interest in connection with a wide range of problems, among them, the Aharonov-Bohm (AB)
effect \cite{ab},  which establishes the reality of the electromagnetic
gauge potential, is a typical example.
In fact, the AB effect forms only the prelude to the even more general class of topological
phenomena which are possible in gauge theories. In particular,
the fractional quantum Hall effect \cite{fqhe1,fqhe2} is  the prominent representative. 
A succinct  account of the latter is given in terms of new
particles called composite fermions (CF) by casting electron-electron correlation in terms
of vortex attachment facility to grasp the  intricate many-particle behavior \cite{jain}.

In the present paper recognizing the significance of the AB non-integrable
phase factor we consider,
inspired by the CF idea, the representation of strongly correlated
electrons as a fermions plus attached ``flux tubes".
This effectively removes most of the electron-electron interaction from the problem
and leads to composite particles which are almost
void of mutual interactions.
Furthermore, taking into account the proper topology of the phase
field dual to the charge, we recognize that
the elementary excitations in strongly correlated system always carry $2\pi$-kinks of the phase
field  characterized by the topological winding number. 
 We reveal the impact
of these  topological excitations  for the  phase diagram of cuprates
 and show that they can induce its
unusual feature: a ``hidden" quantum critical point of a novel type that
is not related to the symmetry breaking.

We consider an effective one--band electronic
Hamiltonian on a tetragonal lattice that emphasize strong anisotropy and
the presence of a layered CuO$_2$ stacking sequence in cuprates:
${\cal H}= {\cal H}_t+{\cal H}_J+{\cal H}_U$, where
\begin{eqnarray}
&&{\cal H}_t=\sum_{\alpha\ell}
\left[ -\sum_{\langle {\bf r}{\bf r}'\rangle}
 (t+\mu\delta_{\bf r,r'})c^{\dagger }_{{\alpha}\ell}({\bf r})
c_{\alpha \ell}({\bf r}')
\right.
\nonumber\\
&&+
\left.
\sum_{{\langle \langle{\bf r}{\bf r}'\rangle\rangle}}
 t'c^{\dagger }_{{\alpha}\ell}({\bf r})
c_{\alpha \ell}({\bf r}')
 -
\sum_{{\bf r}{\bf r}'}
t_\perp({\bf r}{\bf r}')
 c^{\dagger }_{{\alpha}\ell}({\bf r})
c_{\alpha \ell+1}({\bf r}')\right],
\nonumber\\
&&{\cal H}_J=\sum_\ell\sum_{{\langle {\bf r}{\bf r}'\rangle}}
{J}\left[{\bf S}_\ell{({\bf r})}
\cdot{\bf S}_\ell{({\bf r}')}
-\frac{1}{4}{n}_\ell{({\bf r})}{n}_\ell{({\bf r}')}\right],
\nonumber\\
&&{\cal H}_U=\sum_{\ell\bf r}
Un_{\uparrow\ell} ({\bf r}) n_{\downarrow\ell}({\bf r}).
\label{mainham}
\end{eqnarray}
Here, $\langle {\bf r},{\bf r}'\rangle$ and 
$\langle\langle {\bf r},{\bf r}'\rangle\rangle$
identifies  summation
over the nearest-neighbor and next--nearest--neighbor
sites labeled by $1\le {\bf r}\le N$ within the CuO  plane, respectively, with $t,t'$
being the {\it bare} 
hopping integrals,
while $1\le\ell\le N_\perp$ labels copper-oxide layers and $t_\perp$ stands for the
 interlayer coupling.
The operator $c_{\alpha\ell}^\dagger({\bf r})(c_{\alpha\ell}({\bf r}))$
creates (annihilates) an electron of spin $\alpha$ at the lattice site $({\bf r},\ell)$.
Next, $S^a_{\ell}({\bf r})=\sum_{\alpha\beta}c^\dagger_{\alpha\ell}({\bf r})
\sigma_{\alpha\beta}^a c_{\beta\ell}({\bf r})$ ($a=x,y,z$) stands for spin
and ${n}_\ell({{\bf r}})= n_{\uparrow\ell} ({\bf r})+n_{\downarrow\ell}({\bf r})$
 number operators,
respectively, where  $ {n}_{\alpha\ell}({{\bf r}})= c^\dagger_{\alpha\ell}({\bf r})
c_{\alpha\ell}({\bf r})$ and $\mu$ is the chemical potential.
 Subsequently, $U$ is the on--site repulsion Coulomb energy of the order of
bandwidth and $J$ the antiferromagnetic (AF) exchange. The values of $J$ in cuprates
are known to be not
strongly dependent on materials with the magnitude of $0.1-0.16$ eV.
The electronic dispersion is
$\epsilon({\bf k},k_z)=\epsilon_\|({\bf k})+\epsilon_\perp({\bf k},k_z)$,
where the in-plane contribution is
 $\epsilon_\|({\bf k})=-2t\left[\cos(ak_x)+\cos(ak_y)\right]+4t'\cos(ak_x)\cos(ak_y)$
with $t'>0$. Furthermore,
 the $c-$axis dispersion is $\epsilon_\perp({\bf k},k_z)=2t_\perp({\bf k})\cos(ck_z)$,
while $t_\perp({\bf k})=t_\perp\left[\cos(ak_x)-\cos(ak_y)\right]^2$
 as predicted on the basis of  band calculations \cite{ander}.

We write the partition function $Z=\int\left[{\cal D}\bar{c}  {\cal D}\bar{c}
\right]e^{-{\cal S}[\bar{c},c]}$ 
with the action ${\cal S}[\bar{c},c]=\int_0^\beta d\tau[\sum_{\alpha{\bf r}\ell}
 \bar{c}_{\alpha\ell}({\bf r}\tau)\partial_\tau{c}_{\alpha\ell}({\bf r}\tau)+{\cal H}(\tau)]$
using  coherent-state fermionic path integral
over Grassmann fields $c_{\alpha\ell}({\bf r}\tau),{\bar c}_{\alpha\ell}({\bf r}\tau)$
depending on the ``imaginary time" $0\le\tau\le \beta\equiv 1/k_BT$
 with $T$ being the temperature.
The last term in Eq.(\ref{mainham}) we write as
${\cal H}_U(\tau)=U\sum_{{\bf r}\ell}\{({1}/{4}){n_\ell}^2({{\bf r}}\tau)
-\left[{\bf \Omega}_\ell({\bf r}\tau)\cdot{\bf S}_\ell({\bf r}\tau)\right]^2\}$ 
singling out the charge U(1)  and  spin   sector in SU(2)/U(1) coset space, where the unit vector
${\bf \Omega}_\ell({\bf r}\tau)$ sets
varying in space-time  spin quantization axis \cite{weng}. In the following we
 fix our attention on 
U(1) invariant {\it charge }sector, and use Hubbard-Stratonovich transformation
to decouple the Coulomb 
term giving  rise to fluctuating  imaginary  ``voltage"  $iV_\ell({\bf r}\tau)$
conjugate to the  number of charged particles $n_\ell({\bf r}\tau)$.
Furthermore,  we write the field  $V_\ell({\bf r}\tau)$ as a sum of
a static $V_{0\ell}({\bf r})$ and periodic function
$\tilde{V}_\ell({\bf r}\tau)\equiv\tilde{V}_\ell({\bf r}\tau+\beta)$:
$V({\bf r}\tau)=V_0({\bf r})+\tilde{V}({\bf r}\tau)$. Using  Fourier series
$\tilde{V}({\bf r}\tau)=({1}/{\beta})\sum_{n=1}^\infty
[\tilde{V}({\bf r}\omega_n)e^{i\omega_n\tau}+c.c.]$
with $\omega_n=2\pi n/\beta$ ($n=0,\pm1,\pm2$)
being the (Bose) Matsubara frequencies.
Now, we introduce the {\it phase } (or ``flux") field ${\phi}_\ell({\bf r}\tau)$
via the Faraday--type relation
\begin{equation}
\dot{\phi}_\ell({\bf r}\tau)\equiv\frac{\partial\phi_\ell({\bf r}\tau)}
{\partial\tau}=\tilde{V}_\ell({\bf r}\tau),
\label{jos}
\end{equation}
to remove the imaginary term
 $i\int_0^\beta d\tau\dot{\phi}_\ell({\bf r}\tau) n_\ell({\bf r}\tau)
\equiv i\int_0^\beta d\tau\tilde{V}_\ell({\bf r}\tau)n_\ell({\bf r}\tau)$
 for all the Fourier modes
of the $V_\ell({\bf r}\tau)$ field, except for  the zero frequency
by performing the local gauge transformation to the {\it new} fermionic
 variables $f_{\alpha\ell}({\bf r}\tau)$:
\begin{eqnarray}
c_{\alpha\ell}({\bf r}\tau)=
\exp\left[ i\int_0^\tau d\tau' \tilde{V}_\ell({\bf r}\tau')\right]
f_{\alpha\ell}({\bf r}\tau)
\label{compo}
\end{eqnarray}
Thus, as a result of Coulomb correlations the electron acquire a phase shift similar to that
in the 	electric (i.e. scalar) AB effect \cite{ab}; 
Eq.(\ref{compo}) means that an electron 
has a composite  nature made of the fermionic part $f_{\alpha\ell}({\bf r}\tau)$
with the attached ``flux" (or  AB phase) $\exp[i\phi_\ell({\bf r}\tau)]$.
The transformed  action ${\cal S}[\bar{c},c]\rightarrow {\cal S}[{\phi},\bar{f},f]$ then reads
\begin{widetext}
\begin{eqnarray}
{\cal S}[{\phi},\bar{f},f]&=&\sum_{ \ell}
\int_0^\beta d\tau\left\{
\frac{1}{U}\sum_{\bf r}\left[ 
\frac{\partial\phi_\ell({\bf r}\tau)}{\partial\tau}\right]^2
+\frac{2\mu}{U}\sum_{ {\bf r}}
\frac{1}{i}\frac{\partial\phi_\ell({\bf r}\tau)}{\partial\tau}
-\bar{\mu}\sum_{{ \bf r}\alpha}
\bar{f}_{\alpha\ell}({\bf r}\tau)
 f_{\alpha\ell}({\bf r}\tau)\right.
\nonumber\\
&-&\sum_{\langle {\bf r}{\bf r}'\rangle} te^{-i[\phi_\ell({\bf r}\tau)
-\phi_\ell({\bf r}'\tau)]}
 \sum_\alpha\bar{f}_{{\alpha}\ell}({\bf r}\tau)
f_{\alpha \ell}({\bf r}'\tau)
+
\sum_{{\langle\langle  {\bf r}{\bf r}'\rangle\rangle}}t'e^{-i[\phi_\ell({\bf r}\tau)
-\phi_\ell({\bf r}'\tau)]}
\sum_\alpha \bar{f}_{{\alpha}\ell}({\bf r}\tau)
f_{\alpha \ell}({\bf r}'\tau)
\nonumber\\
&-&\left.
\sum_{{\bf rr}' }
t_\perp({\bf r}{\bf r}')e^{-i[\phi_\ell({\bf r}\tau)
-\phi_{\ell+1}({\bf r}'\tau)]}
\sum_\alpha\bar{f}_{{\alpha}\ell}({\bf r}\tau)
f_{\alpha \ell+1}({\bf r}'\tau)-J\sum_{\langle{\bf r r'\rangle}}
\bar{\cal B}_\ell( {\bf r}\tau,{\bf r}'\tau)
{\cal B}_\ell( {\bf r}\tau,{\bf r}'\tau)\right\}
\label{explicit}
\end{eqnarray}
\end{widetext}
where $\bar{\mu}=\mu-n_f{U}/{2}$ and $n_f=\langle \bar{f}_{\alpha}({\bf r}\tau)
f_{\alpha\ell}({\bf r}\tau)\rangle$ is the occupation number for $f-$fermions 
while $\bar{\cal B}_\ell( {\bf r}\tau,{\bf r}'\tau)
=({1}/{\sqrt{2}})[\bar{f}_{\uparrow\ell}({\bf r}\tau)
\bar{f}_{\downarrow\ell}( {\bf r}'\tau)
-\bar{f}_{\downarrow\ell}( {\bf r}\tau)
\bar{f}_{\uparrow\ell}( {\bf r}'\tau)]$ is the singlet pair (valence bond) operator \cite{bond}.
The chief merit of the transformation in Eq.(\ref{explicit}) is that we have managed 
to cast the strongly correlated problem into a system of
{\it weakly} interacting $f$-fermions with residual interaction given by  $J$,
submerged in the bath of strongly fluctuating U(1) gauge potentials
 (on the {\it high}  energy scale set by $U$)
minimally coupled to $f$-fermions via ``dynamical Peierls" phase factors.
It is clear that the action of these phase factors ``frustrates" the motion of the
fermionic subsystem. However, it is only when charge  fluctuations become {\it phase coherent}
the frustration of the kinetic energy is released.  On average, the effect of this frustrated
motion is the effective mass enhancement of  carriers due to the band narrowing, so that
the ``dressed" band parameters 
$t^\star_X=t_X \langle e^{-i[\phi_\ell({\bf r}\tau)-\phi_\ell({\bf r}'\tau)]}\rangle$
(where $t_X=t,t',t_\perp$) are  used to match electronic spectra of HTSC using 
{\it low}--energy scale $t-J$ model \cite{spectra}. Typically, in cuprates
$t^\star\sim 0.5$ eV, $t'^\star/t^\star\sim 0.15-0.35$ and
$t^\star_\perp$ is of order of magnitude
smaller than the in--plane hopping parameters \cite{ander}.

Because for SC system the   {\it charge quantization} matters, the electromagnetic  U(1)
group governing the phase field is {\it compact}, {\it i.e.}
$\phi_\ell({\bf r}\tau)$  has the topology of a circle ($S_1$).
Genuine topological effects can arise due to  non-homotopic 
mappings of the configuration space onto the gauge group $S_1\to$ U(1).
The total time derivative Berry phase \cite{berry} imaginary term
in Eq.(\ref{explicit})  is {\it nonzero} due to
phase field configurations  with
$\phi_\ell({\bf r}\beta)-\phi_\ell({\bf r}0)=2\pi m_\ell({\bf r})$ 
where $m_\ell({\bf r})=0,\pm1,\pm 2,\dots$
marks the U(1) {\it winding } (or Chern) number.
Therefore, the proper  integration measure over $\phi$ in a {\it multiply--connected}
 domain is then \cite{schulman}:
\begin{eqnarray}
\int\left[{{\cal D}\phi }  \right]\dots&&\equiv
\sum_{ \{m_\ell({\bf r})\}}\int_0^{2\pi}\prod_{{\bf r}\ell}d\phi_{0\ell}({\bf r})\times
\nonumber\\
&&\times\int_{\phi_\ell({\bf r}0)=\phi_{0\ell}({\bf r})}^{\phi_\ell({\bf r}\beta)
=\phi_{\ell 0}({\bf r})+2\pi m_\ell({\bf r})}
\prod_{{\bf r}\ell\tau}{d}\phi_\ell({\bf r}\tau)\dots
\label{u1path}
\end{eqnarray}
where in each topological sector the integration goes over the gauge potentials with the
Chern number equal to $m_\ell({\bf r})$.
This is an important observation since these global topological  effects are {\it not} encoded
in the operator algebra of the original
 $c_{\alpha\ell}^\dagger({\bf r}),c_{\alpha\ell}({\bf r})$ operators.

To address the issue of the phase order we trace over
the fermionic degrees of freedom using Eq.(\ref{explicit})
to obtain an effective action  is the phase field:
\begin{eqnarray}
{\cal S}[\phi]&=&\sum_{\ell}\int_0^\beta d\tau \left\{\sum_{ {\bf r}}
\left[ 
\frac{1}{U}\dot{\phi}^2_\ell({\bf r}\tau)
+\frac{2\mu}{U}
\frac{1}{i}
\dot{\phi}_\ell({\bf r}\tau)\right]\right.
\nonumber\\
&-&\sum_{{\langle{\bf r}{\bf r}'\rangle}}
{\cal J}_\|(\Delta)\cos\left[2\phi_\ell({\bf r}\tau)
-2\phi_\ell({\bf r}'\tau)\right]
\nonumber\\
&-&\sum_{{\langle\langle{\bf r}{\bf r}'\rangle\rangle}}
{\cal J}'_\|(\Delta)\cos\left[\phi_\ell({\bf r}\tau)
-\phi_\ell({\bf r}'\tau)\right]
\nonumber\\
&-&\left.\sum_{{\bf r}}
{\cal J}_\perp(\Delta)\cos\left[2\phi_\ell({\bf r}\tau)
-2\phi_{\ell+1}({\bf r}\tau)\right]\right\},
\label{phasemodel}
\end{eqnarray}
where the microscopic {\it phase stiffnesses} to the lowest order in the hopping amplitudes
 are given by
\begin{eqnarray}
{\cal J}_\|(\Delta)&=&\frac{t^2}{4}
\int_{-2}^{2}dxdy\frac{x^2y^2}{y^2-x^2}\rho(x)\rho(y)\times
\nonumber\\
&&\times
\left\{\frac{\tanh\left[\frac{1}{2}\beta \epsilon(x)  \right]}{\epsilon(x)}-
\frac{\tanh\left[\frac{1}{2}\beta \epsilon(y)  \right]}{\epsilon(y)}\right\},
\nonumber\\
{\cal J}'_\|(\Delta)&=&-t'\bar{\mu}\int_{-2}^2 dx
\frac{\bar{\rho}(x)}{\epsilon(x)}
\tanh\left[\displaystyle
\frac{1}{2} \beta\epsilon(x)  \right],
\nonumber\\
{\cal J}_\perp(\Delta)&=&{\frac{9{t}^2_{\perp}|\Delta|^2}{16}}
\int_{-2}^2dx 
\frac{x^2\rho(x)}{\epsilon^{3}(x)}
\left\{2\tanh\left[\frac{\beta \epsilon(x)}{2}  \right]
\right.
\nonumber\\
&&-\left.\beta\epsilon(x)
{\rm sech}^2\left[ \frac{\beta \epsilon(x)}{2}  \right]  \right\}
\label{stiff}
\end{eqnarray}
Here, we denote $\epsilon(x)=\sqrt{ \bar{\mu}^2+|\Delta|^2 x^2}$ and
$\rho(x)=({1}/{\pi^2}){\bf K}(\sqrt{1-({x^2}/{4})} )$ 
while $\bar{\rho}(x)=\rho(x)-
(2/\pi^2){\bf E}( \sqrt{1-({x^2}/{4})})$,
where ${\bf K}(x)$ and  ${\bf E}(x)$ are the complete  elliptic integrals 
of the first and second kind, respectively \cite{EllipticFunction}.
While ${\cal J}_\|$ and ${\cal J}_\perp$ depend on the square
of the corresponding  hopping elements, which render them similar to the
Josephson {\it pair} tunneling  amplitudes, the stiffness ${\cal J}'_\|$
is different: it depends {\it linearly} on $t'$
and governs the process of correlated   {\it particle-hole} motion.
Collective  pair (and in general multiple charge) tunneling events  are costly for large $U$,
so that  excitonic coherent charge transfer dominates the in-plane physics \cite{kopec}.
The inter-plane stiffness ${\cal J}_\perp$  is essential, however, in establishing 
bulk superconductivity via the Josephson coupling. All the stiffnesses in Eq.(\ref{stiff})
rest on a gap due to the  in-plane  {\it momentum space }  pairing of the $f$-fermions 
induced by    AF exchange $J$: a Gorkov-type decoupling 
of the valence bond term in Eq.(\ref{explicit})  readily gives for the
gap parameter $|\Delta({\bf k})|$
\begin{eqnarray}
&&1
=\frac{J}{N}\sum_{\bf k}\frac{[\cos(k_xa)
-\cos(k_ya)]^2}{2{\cal E}({\bf k})}
\tanh\left[
\frac{\beta{\cal E}({\bf k})}{2} \right],
\label{rvbsol}
\end{eqnarray}
with the quasiparticle spectrum of the $f$-fermions,
 ${\cal E}^2({\bf k})=[\epsilon_\|^\star({\bf k})-\bar{\mu}]^2
+|\Delta({\bf k})|^2 $ and  $\Delta({\bf k})=\Delta[\cos(k_xa)
-\cos(k_ya)]$. Obviously, the presence of the ``$d$-wave" pair function $\Delta({\bf k})$
is {\it not} a signature of the superconducting state-it merely marks the region of
non-vanishing phase stiffness.
The state with truly  off-diagonal long range order
is  signalled by $\langle e^{i\phi_\ell({\bf r}\tau)}\rangle\neq 0 $ marking the
{\it macroscopic quantum phase coherence}. To proceed,
we introduce  the unimodular complex scalar
 $z_\ell({\bf r}\tau)=e^{i\phi_\ell({\bf r}\tau)}$ and rewrite the partition function as
$Z=\int\left[{\cal D}^2 {z}\right]\prod_{{\bf r}\ell}
\delta\left(|{z_\ell({\bf r}\tau)}|^2-1\right)e^{-{\cal S}[z,z^\star]}$
where the unimodularity constraint can be imposed
with a real Lagrange multiplier  $\lambda$, so that the effective action reads
\begin{eqnarray}
	{\cal S}[z,z^\star]=
\frac{1}{\beta N N_\perp}\sum_{{\bf q}\omega_n}
  {z}^\star({\bf q},\omega_n)\Gamma^{-1}({\bf q},\omega_n)
 {z}({\bf q},\omega_n).
\end{eqnarray}
Here, ${\bf q}\equiv({\bf k},k_z)$  and 
$\Gamma^{-1}({\bf q},\omega_n)=\lambda-\Sigma({\bf q},\omega_n)
+\gamma^{-1}_{0}(\omega_n)$. Subsequently,
$\Sigma_{\ell\ell'}({\bf r}\tau,{\bf r'}\tau')=
{\cal J}_\perp\delta_{{\bf r},{\bf r}'}\delta_{	{\ell-\ell'},1}+
{\cal J}'_\|\delta_{\ell\ell'}\delta_{{\bf r}-{\bf r}',{\bf d}_{2nd}}
+\bar{\cal J}_\|\Gamma_{\ell\ell'}({\bf r}\tau,{\bf r'}\tau')
\delta_{{\bf r}-{\bf r}',{\bf d}_{1st}}$ while ${\bf d}_{1st}$ and
${\bf d}_{2nd}$ are the lattice vectors connecting nearest neighbors
and next-nearest neighbors in the CuO plane, respectively.
Furthermore, $\gamma_{0}(\omega_n)$ is the Fourier transform of the
bare phase correlator 
$\langle e^{-i[\phi_\ell({\bf r}\tau)-\phi_{\ell'}({\bf r}'\tau')]}\rangle_0$
originating from the kinetic and  topological part of the action in Eq.(\ref{phasemodel}).
Using Poisson summation formula we obtain
\begin{eqnarray}
&&\langle e^{-i[\phi_\ell({\bf r}\tau)-\phi_{\ell'}({\bf r}'\tau')]}\rangle_0=
\frac{\vartheta_3\left(\frac{2\pi\mu}{U}+\pi\frac{\tau-\tau'}{\beta},
e^{-\frac{4\pi^2}{\beta U}}  \right)}{\vartheta_3\left(\frac{2\pi\mu}{U},
e^{-\frac{4\pi^2}{\beta U}}  \right)}
\nonumber\\
&&\times\exp\left\{-{U}\left[|\tau-\tau'|-\frac{(\tau-\tau')^2}{\beta} \right]  \right\}
\delta_{\bf r,r'}\delta_{\ell\ell'}
\end{eqnarray}
where $\vartheta_3(z,q)$ is the Jacobi theta function \cite{EllipticFunction}
which is $\beta$-periodic in the
imaginary-time $\tau$ as well as in the variable $2\mu/U$ with the period of unity
which emphasize the special role of the integer values of the reduced chemical potential.
At criticality, the condition  $\Gamma^{-1}({\bf 0},0)|_{\lambda=\lambda_c}=0$ fixes the Lagrange
parameter while the phase-coherence boundary  is:
$1=\left.\frac{1}{\beta N N_\perp}\sum_{{\bf q}\omega_n}
{ \Gamma}({\bf q},\omega_n)\right|_{\lambda=\lambda_c}$.
Theoretically it is much simpler to consider the fixed chemical
potential case and the
results at fixed density can always be obtained after a Legendre
transformation.
The resulting temperature-chemical potential phase diagram is depicted in Fig.1.
 First, it  shows that
$T_c$  correlates with the diagonal hopping $t'$ in accord with the observation 
that the next-nearest-neighboring hopping  dominates the variation of
the maximum $T_c$ in hole doped cuprates \cite{pavarini}. Further, the phase diagram
in Fig.1 exhibits the special point at
 $\mu_{c}$ defined by $2\mu_c/U=1/2$ away from the incompressible  Mott state at $2\mu_c/U=1$
from which the superconducting lobe emanates.
\begin{figure}
\begin{center}
\includegraphics[width=8cm]{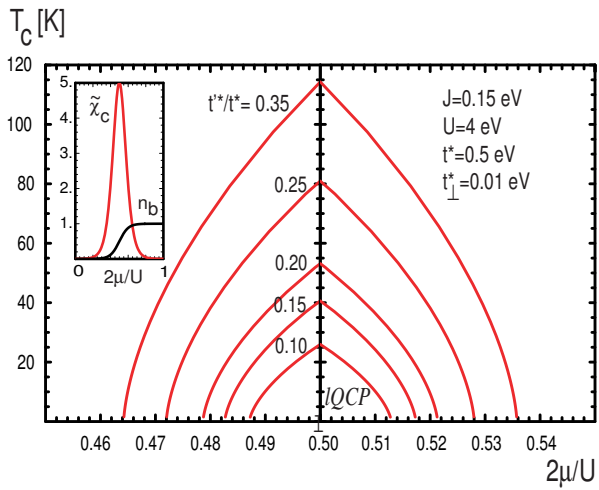}
\end{center}
 \caption{(Color online) The critical temperature $T_c$ as a function of the 
chemical potential  for the parameters as indicated in the plot.
The label  lQCP marks the special point inside the superconducting lobe where the local charge susceptibility
diverges and    superconductivity is most robust.
Inset: $\tilde {\chi}_c\equiv U\chi_c/2$ and the occupation number $n_b$ for $T=0.1U$.
   }\label{fig1}
\end{figure} 
In cuprates there is clear evidence for the existence of a special
doping point $x_c$ in the lightly-overdoped region  where superconductivity is
most robust. Such behavior indicates this point could be a 
QCP while  the associated critical fluctuations might be responsible for the 
unconventional normal state behavior \cite{qcp}. However, the resemblance
to a conventional QCP is {\it incomplete} due to the lack of any clear signature of
thermodynamic critical behavior: a QCP is generally the
end-point of a line of phase transition.
Experiments appear to exclude any broken  symmetry 
around this point  although  a sharp change in transport properties is 
observed \cite{transport} and $\partial\mu/\partial x$
becomes vanishingly small due to slow chemical potential shift implying a divergence of
charge susceptibility \cite{chemical}.
We argue  that due topological excitations  indeed such a singularity   arises at $\mu_c$ in the 
{\it local}  charge susceptibility  $\chi_c=\partial n_e/\partial\mu$ where
 $n_e\equiv\langle \bar{c}_{\alpha}({\bf r}\tau)c_{\alpha\ell}({\bf r}\tau)\rangle$ 
is the electron filling. From Eq.(\ref{explicit}) we readily obtain that
$n_e=n_f+ n_b-{2\mu}/{U}$ where the topological contribution is 
given by $n_b={2\mu}/{U}+({2}/{iU})
\langle\dot{\phi}({\bf r}\tau)\rangle$. In the large--$U$ limit 
$\mu\to n_fU/2$ so that $n_e\to n_b$, which means,
via Eq.(\ref{u1path}), that  for strong correlations $n_e$ is governed
 by the topological winding numbers rather then  the
number of fermionic oscillators.
 However, the winding  number
 is a topologically conserved quantity
and is ``protected"  against
the small changes of $\mu$. Being 	an integer it can not change 
at all if it has to change continuously.
For  substantial perturbations the  ground state crosses over abruptly to other eigenstates:
$n_b$ can change only when level degeneracies occur which happens at  isolated discrete values of
$2\mu/U$. For $T=0$, specializing to the topological and the leading kinetic term in Eq.(\ref{phasemodel}) we obtain
$n_b=2\mu/U-h(2\mu/U)$ where $h(\xi+1/2)+1/2=\xi-[\xi]$, while $[\xi]$ is 
the greatest integer less than or equal to $\xi$. Clearly,  charge susceptibility diverges at
$\mu_c$, thus marking the {\it local}  QCP of a novel  type, not related to the paradigm of symmetry breaking but
resulting from topological effects in strongly correlated system.
	This is substantiates the notion of the
	topological order \cite{wen}, which differentiate the electronic ground state
	into two states  labelled by the topological winding number
	and with the degeneracy point separating  them. It
	controls a remarkable concurrence between normal state
	properties and the ground-state superconducting
	order  setting up a unique critical doping
	point in the phase diagram where the transport properties change
	very suddenly and where superconductivity is most robust.
It is also an example of  the emergent phenomenon since its  topological underpinning
(which has to be recognized) is not simply encoded  in the microscopic
 electronic model, Eq.(\ref{mainham}), in which it arises.

This work was supported by the Polish Science Committee (KBN)
under the grant No. 2P03B 009 25 in years 2003-2005.


 \end{document}